\documentclass[12pt]{article}
\usepackage{amsfonts,bm}
\usepackage{graphicx}
\baselineskip
\offinterlineskip
\newcommand{\p}{\partial}
\begin{document}

\begin{center}
{\bf Hyperdeterminant and an integrable partial differential equation}
\end{center}

\begin{center}
{\bf Willi-Hans Steeb} \\[2ex]
International School for Scientific Computing, \\
University of Johannesburg, Auckland Park 2006, South Africa, \\
e-mail: {\tt steebwilli@gmail.com}
\end{center}

\strut \hfill

{\bf Abstract.} We discuss an integrable partial differential equation
arising from the hyperdeterminant.
\newline

\newline

It is well known [1,2,3,4] that the Bateman equation in two dimensions
$$
\frac{\p^2 u}{\p x_1^2}\left(\frac{\p u}{\p x_2}\right)^2 
+ \frac{\p^2 u}{\p x_2^2}\left(\frac{\p u}{\p x_1}\right)^2 
-2\frac{\p u}{\p x_1}\frac{\p u}{\p x_2}\frac{\p^2 u}{\p x_1\p x_2} = 0
$$
can be derived from the condition
$$
\det \pmatrix { 0 & \p u/\p x_1 & \p u/\p x_2 \cr
\p u/\p x_1 & \p^2 u/\p x_1^2 & \p^2 u/\p x_1\p x_2 \cr
\p u/\p x_2 & \p^2 u/\p x_1\p x_2 & \p^2 u/\p x_2^2 } = 0 \,.
$$
It is also well known (see for example [5,6]) that the Bateman equation 
can be linearized by a Legendre transformation and the general implicit
solution is given by
$$
x_1 f(u(x_1,x_2)) + x_2 g(u(x_1,x_2)) = c
$$
where $f$ and $g$ are smooth functions.
This partial differential equation plays a central role in studying
the integrability of partial differential equation using the Painlev\'e
test [5,7]. For example the differential equation appears in the Painleve 
analysis of the inviscid Burgers equation, double sine-Gordan,
discrete Boltzmann equation. 
\newline

Here we generalize the condition given above from 
the determinant to the $2 \times 2 \times 2$ hyperdeterminant
of a $2 \times 2 \times 2$ hypermatrix and derive the nonlinear
partial differential equation and discuss its properties.
The extension to $2 \times 2 \times 2 \times 2$ hyperdeterminants
will be straightforward. 
\newline

Cayley [8] in 1845 introduced the hyperdeterminant. Gelfand et al [9]
give an in debt discussion of the hyperdeterminant.
The hyperdeterminant arises as entanglement measure for three qubits
[10,11,12], in black hole entropy [13].
The Nambu-Goto action in string theory can be expressed in terms
of the hyperdeterminant [14]. A computer algebra program for
the hyperdeterminant is given by Steeb and Hardy [11] 
\newline

Let $\epsilon_{00}=\epsilon_{11} = 0$, 
$\epsilon_{01}=1$, $\epsilon_{10}=-1$, 
i.e. we consider the $2 \times 2$ matrix 
$$
\epsilon = \pmatrix { 0 & 1 \cr -1 & 0 } \,. 
$$
Then the determinant of a $2 \times 2$ matrix
$A_2=(a_{ij})$ with $i,j=0,1$ can be defined as
$$
\det A_2 := \frac12 \sum_{i=0}^1 \sum_{j=0}^1 \sum_{\ell=0}^1
\sum_{m=0}^1 \epsilon_{ij} \epsilon_{\ell m} a_{i\ell} a_{jm} \,.
$$
Thus $\det A_2=a_{00}a_{11}-a_{01}a_{10}$.
In analogy the hyperdeterminant of the $2 \times 2 \times 2$ array 
$A_3=(a_{ijk})$ with $i,j,k=0,1$ is defined as
$$
\mbox{Det} A_3 := -\frac12 
\sum_{ii'=0}^1 \sum_{jj'=0}^1 \sum_{kk'=0}^1 \sum_{mm'=0}^1 \sum_{nn'=0}^1 
\sum_{pp'=0}^1 
\epsilon_{ii'} \epsilon_{jj'} \epsilon_{kk'} \epsilon_{mm'} \epsilon_{nn'} 
\epsilon_{pp'} 
a_{ijk} a_{i'j'm} a_{npk'} a_{n'p'm'} \,.  
$$
There are $2^8=256$ terms, but only 24 are nonzero. We find
\begin{eqnarray*}
\mbox{Det} A_3 &=& a_{000}^2 a_{111}^2 + a_{001}^2 a_{110}^2 + a_{001}^2
a_{101}^2 + a_{100} a_{011}^2 \cr
&& -2(a_{000} a_{001} a_{110} a_{111} + a_{000} a_{010} a_{101} a_{111} \cr
&& + a_{000} a_{100} a_{011} a_{111} + a_{001} a_{010} a_{101} a_{110} \cr
&& + a_{001} a_{100} a_{011} a_{110} + a_{010} a_{100} a_{011} a_{101}) \cr
&& + 4(a_{000} a_{011} a_{101} a_{110} + a_{001} a_{010} a_{100} a_{111}) \,.
\end{eqnarray*}

The hyperdeterminant $\mbox{Det}(A)$ of the three-dimensional array 
$A_3=(a_{ijk}) \in {\mathbb R}^{2 \times 2 \times 2}$
can also be calculated as follows
\begin{eqnarray*}
\mbox{Det}(A_3) &=& \frac14\left( 
\det\left(\pmatrix { a_{000} & a_{010} \cr a_{001} & a_{011} } + 
\pmatrix { a_{100} & a_{110} \cr a_{101} & a_{111} }\right)\right. \\
&& \left. -\det\left( \pmatrix { a_{000} & a_{010} \cr a_{001} & a_{011} } 
- \pmatrix { a_{100} & a_{110} \cr a_{101} & a_{111} } \right) \right)^2 \\
&& -4 \det\pmatrix { a_{000} & a_{010} \cr a_{001} & a_{011} } 
\det\pmatrix { a_{100} & a_{110} \cr a_{101} & a_{111} } \,.
\end{eqnarray*}
If only one of the coefficients $a_{ijk}$ is nonzero
we find that the hyperdeterminant of $A_3$ is 0.
\newline

Given a $2 \times 2 \times 2$ hypermatrix $A_3=(a_{jk\ell})$, 
$j,k,\ell=0,1$ and the $2 \times 2$ matrix 
$$
S = \pmatrix { s_{00} & s_{01} \cr s_{10} & s_{11} } \,.
$$
The multiplication $A_3S$ which is again a $2 \times 2$ hypermatrix
is defined by 
$$
(A_3S)_{jk\ell} := \sum_{r=0}^1 a_{jkr}s_{r\ell} \,.
$$
If $\det(S)=1$, i.e. $S \in SL(2,{\mathbb C})$, then
$\mbox{Det}(A_3S)=\mbox{Det}(A_3)$. 
\newline

In analogy with the Bateman equation we set
$$
a_{000} \mapsto 0, \quad a_{001} \mapsto \frac{\p u}{\p x_3}, \quad
\quad a_{010} \mapsto \frac{\p u}{\p x_2}, \quad
\quad a_{100} \mapsto \frac{\p u}{\p x_1},
$$
$$
a_{011} \mapsto \frac{\p^2 u}{\p x_2\p x_3}, \quad
a_{101} \mapsto \frac{\p^2 u}{\p x_1\p x_3}, \quad
a_{110} \mapsto \frac{\p^2 u}{\p x_1\p x_2}, \quad
a_{111} \mapsto \frac{\p^3 u}{\p x_1 \p x_2 \p x_3}
$$
and obtain the nonlinear partial differential equation
\begin{eqnarray*}
&& \!\!\!\left(\frac{\p u}{\p x_1}\right)^2
\left(\frac{\p^2 u}{\p x_2\p x_3}\right)^2 +
\left(\frac{\p u}{\p x_2}\right)^2 
\left(\frac{\p^2 u}{\p x_1 \p x_3}\right)^2 +
\left(\frac{\p u}{\p x_3}\right)^2
\left(\frac{\p^2 u}{\p x_1\p x_2}\right)^2 \\ 
&& \!\!\!-2\left(
\frac{\p u}{\p x_1}\frac{\p u}{\p x_2}
\frac{\p^2 u}{\p x_1\p x_3}\frac{\p^2 u}{\p x_2\p x_3} +
\frac{\p u}{\p x_1}\frac{\p u}{\p x_3}
\frac{\p^2 u}{\p x_1\p x_2}\frac{\p^2 u}{\p x_2\p x_3} +
\frac{\p u}{\p x_2}\frac{\p u}{\p x_3}
\frac{\p^2 u}{\p x_1\p x_2}\frac{\p^2 u}{\p x_1\p x_3}\right) \\
&& \!\!\!+ 4\frac{\p u}{\p x_1}\frac{\p u}{\p x_2}\frac{\p u}{\p x_3}
\frac{\p^3 u}{\p x_1\p x_2 \p x_3} = 0 \,. 
\end{eqnarray*}
The partial differential equation is invariant under the permutations
of $x_1,x_2,x_3$. The group of symmetries is $SL(3,{\mathbb R})$.
The equation can also be linearized by a Legendre transformation.

\strut\hfill

[1] Fairlie D. B., Govaerts J. and Morzov A.,
Universal field equations and covariant solutions,
{\it Nuclear Phys. B} {\bf 373} (1992) 214-232.
\newline

[2] Fairlie D. B., Integrable systems in higher dimensions,
{\it Prog. Theor. Phys. Supp.} 1995, N 118, 309-327.
\newline  

[3] Derjagin V. and Leznov A., Geometrical symmetries of the
universal equation, {\it Nonlinear Mathematical Physics},
{\bf 2} (1995), 46-50. 
\newline

[4] Euler N., Lindblom O., Euler M. and Persson L.-E.,
The higher dimensional Bateman equation and Painlev\'e 
analysis of nonintegrable wave equations, 
{\it Symmetry in Nonlinear Mathematical Physics}, {\bf 1} (1997),
185-192.
\newline

[5] Weiss J., The sine-Gordon equations: Complete and partial 
integrability, {\it J. Math. Phys.} {\bf 25} (1984) 226-2235. 
\newline

[6] Steeb W.-H., Problems and Solution in Theoretical and
Mathematical Physics, Volume I,
3rd edition, World Scientific, Singapore (2010).  
\newline

[7] Steeb W.-H. and Euler N., 
Nonlinear Evolution Equations and Painlev\'e Test, 
World Scientific, Singapore (1988).
\newline

[8] Cayley A., On the theory of linear transformations, 
{\it Camb. Math. J.} {\bf 4} (1845) 193-209. 
\newline

[9] Gelfand I. M., Kapranov M. M. and Zelevinsky A.,
Discriminants, Resultants, and Multidimensional Determinants,
Birkh\"auser, Basel (2008).
\newline

[10] Coffman V., Kundu J. and Wooters W.,
Distributed Entanglement, {\it Phys. Rev. A} {\bf 61} 052306 (2000).
\newline

[11] Steeb W.-H. and Hardy Y.,
Quantum Mechanics using Computer Algebra,
World Scientific, Singapore (2010). 
\newline

[12] Steeb W.-H. and Hardy Y.,
Problems and Solutions in Quantum Computing and Quantum Information,
3rd edition, World Scientific, Singapore (2011).
\newline

[13] Duff M., String triality, black hole entropy, and Cayley's
hyperdeterminant, {\it Phys. Rev. D} {\bf 76} (2007) 025017.

[14] Hitoshi Nishino and Subhash Raipoot,
Nambu-Goto actions and Cayley's hyperdeterminant,
{\it Phys. Lett. B} {\bf 652} (2007), 135-140.

\end{document}